\documentclass[11pt,prd,amsmath,amsfonts,amssymb,notitlepage,nofootinbib,a4paper]{revtex4-1}
\usepackage{enumerate}
\usepackage{amsthm}
\usepackage{graphicx}

\begin{document}

\title{Causal structure  and electrodynamics on Finsler spacetimes}
\author{Christian Pfeifer}
\email{christian.pfeifer@desy.de}
\affiliation{II. Institut f\"ur Theoretische Physik und Zentrum f\"ur Mathematische Physik, Universit\"at Hamburg, Luruper Chaussee 149, 22761 Hamburg, Germany}
\author{Mattias N.\,R. Wohlfarth}
\email{mattias.wohlfarth@desy.de}
\affiliation{II. Institut f\"ur Theoretische Physik und Zentrum f\"ur Mathematische Physik, Universit\"at Hamburg, Luruper Chaussee 149, 22761 Hamburg, Germany}

\begin{abstract}
We present a concise new definition of Finsler spacetimes that generalize Lorentzian metric manifolds and provide consistent backgrounds for physics. Extending standard mathematical constructions known from Finsler spaces we show that geometric objects like the Cartan non-linear connection and its curvature are well-defined almost everywhere on Finsler spacetimes, also on their null structure. This allows us to describe the complete causal structure in terms of timelike and null curves; these are essential to model physical observers and the propagation of light. We prove that the timelike directions form an open convex cone with null boundary as is the case in Lorentzian geometry. Moreover, we develop action integrals for physical field theories on Finsler spacetimes, and tools to deduce the corresponding equations of motion. These are applied to construct a theory of electrodynamics that confirms the claimed propagation of light along Finsler null geodesics.
\end{abstract}
\maketitle

\newcommand{\new}[1]{$\blacktriangleright$#1$\blacktriangleleft$}

\section{Motivation}\label{sec:mot}
Dynamical physical processes could not be described without using clocks. Independently of how precisely clocks are realized in experiment, it is absolutely essential to model them in theory. In relativistic theories this is facilitated by a geometric clock postulate that provides the proper time along the worldline $\tau\mapsto x(\tau)$ of an observer, or massive particle, moving through the spacetime manifold~$M$. The most general geometric clock postulate for which proper time~$T[x]$ depends locally on the position and four-velocity is a reparametrization-invariant functional of the form
\begin{equation}\label{eqn:Flength}
T[x]=\int d\tau \, F(x(\tau),\dot x(\tau))\,.
\end{equation}
For time measurements in general relativity the function~$F$ is determined by a Lorentzian metric $\tilde g_{ab}$ as $F(x,\dot x) = | \tilde g_{ab}(x)\dot x^a \dot x^b |^{1/2}$.  More general $F$ are the basic ingredient of Finsler geometry~\cite{Asanov,Chern,Bucataru}, which generalizes Riemannian geometry so that all geometric objects and fields depend not only on the points of the manifold but also on its tangent directions. 

In physics, the integral~$T[x]$ in (\ref{eqn:Flength}) plays a fundamental role. On the one hand, it provides a definition of proper time that is independent of the actual clock used; on the other hand, the worldlines of observers and massive particles are determined as the curves that extremize the integral. 

Because of the generality of the Finsler clock postulate it is no surprise that Finsler geometry has emerged in a number of different physical contexts in recent years. From the perspective of fundamental physical theory the importance of Finsler geometry was realized: in various approaches to quantum gravity to describe the effective classical geometry of Planck scale modified dispersion relations and the breaking of local Lorentz invariance, e.g. in~\cite{AmelinoCamelia:1996pj,Gambini:1998it,Girelli:2006fw,Gibbons:2007iu}; for generalized backgrounds defined by hyperbolic polynomials~\cite{Raetzel:2010je}; and for covariant formulations of electrodynamics in very general linear and non-linear optical media~\cite{Hehl:2002hr,Schuller:2009hn,Visser:2002sf}. On a more phenomenological level, simple Finsler backgrounds allow fits of astronomical and cosmological data that are not satisfactorily explained by general relativity: the Pioneer anomaly, dark matter and dark energy~\cite{Chang:2008yv,Chang:2009pa,Li:2009sv}.

This appears very promising; a closer inspection, however, reveals that it is not at all straightforward to describe the fundamental geometric structure of spacetime by Finsler geometry. In order to do so a number of basic conceptual issues need to be resolved first:
\begin{itemize}
\item The mathematical foundations, including the construction of connections, covariant derivatives and curvature, 
are tailored for the case of definite Finsler metrics that are a direct generalization of definite Riemannian geometry. These constructions are not directly applicable to generalize the Lorentzian signature case. Previous work to remedy this situation turns out not to be sufficient: Finsler metrics with globally Lorentzian signature, e.g. in~\cite{Bucataru}, do not include the metric limit; in certain Finsler spaces~\cite{Asanov} it is not possible to discuss the motion of light; other definitions~\cite{Beem} exclude too many interesting applications like that to electrodynamics in optical media. 
\item The causal structure of a Finsler spacetime needs to be clarified, as has been observed in~\cite{Beem,Perlick:2005hz}. A good definition must provide a precise notion of timelike vectors, and hence a clear definition of observers. At the same time it must allow the discussion of null motion which is essential to describe the effective propagation of light. 
\item Action principles for the formulation of field theories on Finsler spacetime should be available. Previous work on this topic includes the osculating formalism~\cite{Asanov} which, however, does not allow the full reconstruction of the
fields from solutions of the equations of motion. Other approaches use actions with divergency problems~{\cite{Voicu:2009wi,Voicu:2010kw}} that technically are not suited to derive equations of motion, or are not immediately related to standard Finsler geometry formulations over the manifold and its tangent bundle~\cite{Vacaru:2003uk}. 
\end{itemize}
In this article we will develop solutions for these conceptual problems. Through our results it becomes possible to use Finsler spacetimes, instead of Lorentzian manifolds, as consistent generalized geometric backgrounds for physics.

In section~\ref{sec:spacetime} we will provide a concise new definition of Lorentzian Finsler spacetimes, along with a discussion of the immediate physical implications. The standard mathematical technology used in Finsler geometry is reviewed in section~\ref{sec:geometry}, before we will prove the important result that Finsler spacetimes in our sense allow a clean extension of the definition of connections, covariant derivatives and curvature to their null structure. In section~\ref{sec:causal} we will demonstrate the existence of open convex cones of timelike vectors with null boundaries as in Lorentzian metric geometry. We will then employ these results to describe the motion of observers and the effective motion of light by means of timelike and null Finsler geodesics. 
The impact of our definition and the proven theorems on the causal structure will be illustrated by means of  two examples in section~\ref{sec:examples}. The first example discusses standard Lorentzian metric spacetimes as a special case of Finsler spacetimes. The second example goes beyond metric geometry; it has a more complicated null structure with two cones of light propagation and is relevant for the description of birefringent optical media.
In section~\ref{sec:actions} we will present a new method for the formulation of well-defined action integrals on Finsler spacetime. This method is based on the restriction of the tangent bundle to a subbundle on which homogeneous Lagrangians can be integrated. With this formalism we will explicitly construct a generalized theory of electrodynamics on Finsler spacetimes in section~\ref{sec:ED}. We will study the propagation of singularities through the corresponding partial differential equations. As a result we can justify for the first time that light in a Finsler spacetime indeed propagates along  Finsler null geodesics. We will conclude with a discussion and an outlook in section~\ref{sec:disc}.

\section{Finsler spacetimes}\label{sec:spacetime}
As discussed above, the description of spacetime by Finsler geometry is not straightforward and poses a number of open questions. In this section we will propose a new definition for Finsler spacetimes to answer these. We will comment on the physical motivation and the immediate consequences of our definition, and on the  interpretation of physical fields on Finsler spacetime. Then, in section~\ref{sec:geometry}, we will introduce
the mathematics of the Cartan non-linear connection and curvature, before we exhibit the causal structure of Finsler spacetimes in section~\ref{sec:causal}.

\subsection{Definition and properties}

Recall that the tangent bundle $TM=\cup_{p\in M}T_pM$ of a four-dimensional manifold $M$ is the union of all its tangent spaces, and is a fibre bundle of dimension eight over $M$. Coordinates $x$ on~$M$ induce coordinates $(x,y)$ on $TM$. Any point $Y\in TM$ is a
tangent vector in some $T_pM$ where $p$ has coordinates $(x^a)$, and can be expressed as $Y = y^a \frac{\partial}{\partial x^a}{}_{|x}$. The
induced coordinates of $Y$ are then defined by~$(x^a,y^a)$. For simplicity we will use induced coordinates throughout this article. The
coordinate basis of $TTM$ will be written with the short-hand notation
\begin{equation}
\Big\{\partial_a=\frac{\partial}{\partial x^a},\bar\partial_a=\frac{\partial}{\partial y^a}\Big\}.
\end{equation}

We will now state our new definition for Finsler spacetimes before commenting on the details, and discussing its advantages in comparison to various known formulations. 

\vspace{6pt}\noindent\textbf{Definition.}
\textit{A Finsler spacetime $(M,L,F)$ is a four-dimensional, connected, Hausdorff, paracompact, smooth manifold~$M$ equipped with a continuous function $L:TM\rightarrow\mathbb{R}$ on the tangent bundle which has the following properties:
\begin{enumerate}[(a)]
\item $L$ is smooth on the tangent bundle without the zero section $TM\setminus\{0\}$;\vspace{-6pt}
\item $L$ is positively homogeneous of real degree $r\ge 2$ with respect to the fibre coordinates of $TM$,
\begin{equation}\label{eqn:hom}
L(x,\lambda y)  = \lambda^r L(x,y) \quad \forall \lambda>0\,;
\end{equation}
\item \vspace{-6pt}$L$ is reversible in the sense 
\begin{equation}\label{eqn:rev} 
|L(x,-y)|=|L(x,y)|\,;
\end{equation}
\item \vspace{-6pt}the Hessian $g^L_{ab}$ of $L$ with respect to the fibre coordinates is non-degenerate on $TM\setminus A$ where~$A$ has measure zero and does not contain the set $\{(x,y)\in TM\,|\,L(x,y)=0\}$,
\begin{equation}
g^L_{ab}(x,y) = \frac{1}{2}\bar\partial_a\bar\partial_b L\,;
\end{equation}
\item \vspace{-6pt}the unit timelike condition holds, i.e., for all $x\in M$ the set 
\begin{equation}
\Omega_x=\Big\{y\in T_xM\,\Big|\, |L(x,y)|=1\,,\;g^L_{ab}(x,y)\textrm{ has signature
}(\epsilon,-\epsilon,-\epsilon,-\epsilon)\,,\, \epsilon=\frac{|L(x,y)|}{L(x,y)}\Big\}
\end{equation}
contains a non-empty closed connected component $S_x\subset \Omega_x\subset T_xM$.
\end{enumerate}
The Finsler function associated to $L$ is defined as $F(x,y) = |L(x,y)|^{1/r}$.}

\vspace{6pt}The Finsler function $F$ describes important physical aspects of Finsler spacetimes $(M,L,F)$ via the length integral~(\ref{eqn:Flength}). We will now derive certain consequences for $F$ from the definition above. Then we will be in the position to discuss the physical properties of Finsler spacetimes.

\vspace{6pt}\noindent\textbf{Corollary.}
\textit{The Finsler function $F$ of a Finsler spacetime $(M,L,F)$ is a continuous function $F:TM\rightarrow\mathbb{R}$ and satisfies:
\begin{enumerate}[(A)]
\item $F$ is smooth on the tangent bundle where $L$ is non-vanishing, $TM\setminus \{L=0\}$;
\item $F$ is non-negative, $F(x,y)\geq 0$;
\item $F$ is positively homogeneous of degree one in the fibre coordinates and reversible, i.e.,
\begin{equation}\label{eqn:abs}
F(x,\lambda y)=|\lambda| F(x,y)\quad\forall \lambda\in\mathbb{R}\,;
\end{equation}
\item the Finsler metric of $F$,
\begin{equation}
g_{ab}(x,y) = \frac{1}{2} \bar\partial_a\bar\partial_b F^2\,,
\end{equation}
is defined on $TM\setminus \{L=0\}$, and is non-degenerate on $TM\setminus (A\cup \{L=0\})$.
\end{enumerate}}

\vspace{3pt}\noindent\textit{Proof.} (A) and (B) immediately follow from the definition $F = |L|^{1/r}$ and property (a) of Finsler spacetimes.  We remark that $F=|L|^{1/r}$ is non-differentiable at all tangent bundle points where~$L$ changes sign. (C) is a consequence of properties (b) and~(c). 

(Observe that we do not have to require absolute homogeneity $L(x,\lambda y)=|\lambda|^rL(x,y)$ for $L$ in order to obtain this property for $F$, but only the weaker assumptions (b), (c); this in principle allows a larger class of Finsler spacetimes: for instance, $L$ could be a homogeneous polynomial of degree three in $y$.) 

The domain of definition of the Finsler metric defined in~(D) is $TM\setminus \{L=0\}$ because this is where derivatives of~$F$ are defined. To see where $g_{ab}$ is non-degenerate we observe that it is related to the Hessian $g^L_{ab}$ and its inverse by the formulae
\begin{equation}\label{eqn:ggL}
g_{ab} = \frac{2 |L|^{2/r}}{rL}\Big(g^L_{ab}+\frac{(2-r)}{2rL}\bar\partial_aL\bar\partial_bL\Big),\qquad
g^{ab} = \frac{rL}{2|L|^{2/r}}\Big(g^{L\,ab}-\frac{2(2-r)}{r(r-1)L}y^ay^b\Big).
\end{equation}
Using the homogeneity properties of $L$, the determinant of the Finsler metric can be calculated as
\begin{equation}
\textrm{det }g = \frac{16|L|^{8/r-4}}{r^4(r-1)}\textrm{det }g^L\,,
\end{equation}
hence the Finsler metric is non-degenerate on $TM\setminus (A\cup \{L=0\})$. $\square$

\vspace{6pt} From the physical point of view Finsler spacetimes guarantee the following desirable features:

A geometric clock postulate along observers' worldlines can be defined by the Finsler length integral~(\ref{eqn:Flength}). This proper time measurement will be positive because of~(B). Moreover, the Finsler length is reparametrization-invariant due to~(C), which ensures that proper time is an intrinsic geometric quantity. 

Finsler spacetimes are symmetric under time-reversal because proper time is in particular invariant under reparametrizations with orientation reversal. To see this, consider a worldline $\tau\mapsto x(\tau)$ in $M$ and a reparametrization $\sigma(\tau)$ with $d\sigma / d\tau<0$; then
\begin{equation}
\int_{\tau_1}^{\tau_2} d\tau \,F(x(\tau),\dot x(\tau)) = -\int_{\sigma(\tau_2)}^{\sigma(\tau_1)} d\sigma \Big(\textrm{sign} \frac{d\tau}{d\sigma}\Big) F(x(\sigma),x'(\sigma))=\int_{\sigma(\tau_2)}^{\sigma(\tau_1)} d\sigma\, F(x(\sigma),x'(\sigma))\,.
\end{equation}
So the proper time of a curve is indeed independent of the choice of orientation. 

We will see in section~\ref{sec:geometry} that we can use $F$ to construct a connection, covariant derivative and the associated curvature on $TM\setminus (A\cup \{L=0\})$. This construction uses the standard mathematical tools of Finsler geometry which are based on (A) and (D). 

The relevance of $L$ and $g^L_{ab}$ in the definition of Finsler spacetimes $(M,L,F)$ is that they allow us to extend all necessary geometric objects to the larger domain $TM\setminus A$. This extension covers the complete null structure of spacetime, which is of particular importance for physics:

All types of null and non-null Finsler geodesics can be discussed on a Finsler spacetime, so that we may obtain a clear understanding of its causal structure.

As part of the causal structure, Finsler spacetimes admit a clear definition of timelike vectors. These are needed as tangents of observer's worldlines. Their existence at every point $x\in M$ is guaranteed by the unit timelike condition~(e) that provides a shell $S_x$ of unit timelike vectors. Indeed, $y\in S_x$ is normalized by $|L(x,y)|=1$ and is timelike with respect to the metric $g^L_{ab}$. This follows from\footnote{Equation (\ref{eqn:gLL}) is a consequence of Euler's theorem  for $n$-homogeneous functions $f(y)$. This theorem is used in several calculations in this article, and states that $y^a\bar\partial_a f(y)= n f(y)$.}
\begin{equation}\label{eqn:gLL}
g^L_{ab}(x,y)y^ay^b=\frac{1}{2}r(r-1) L(x,y)
\end{equation}
which implies $\textrm{sign}(g^L_{ab}(x,y)y^ay^b)=L/|L|=\epsilon$. On $S_x$,  the signature of $g^L_{ab}$ is
$(\epsilon,-\epsilon,-\epsilon,-\epsilon)$ and so the sign $\epsilon$ indeed defines the timelike direction.  
In section~\ref{sec:causal} we will prove that the unit timelike condition implies that the timelike observer directions on a Finsler spacetime form open convex cones.

In section \ref{sec:examples} we will demonstrate in detail that a Lorentzian metric manifold satisfies all requirements of our definition, and so is a special case of a Finsler spacetime. We will see that it is not possible to model such a manifold using a Finsler metric of globally constant signature, as prescribed in earlier definitions, e.g. in~\cite{Beem,Bucataru}. In the same section we will also discuss a more complicated example of a Finsler spacetime where the full power of our new definition comes into play. This example could be of relevance for a covariant spacetime description for crystal optics; it has $r=4$ and multiple signature changes in the Finsler metric. Again this is excluded by earlier definitions. Finsler spacetimes in our sense are generalizations of Lorentzian spacetimes.

It is important to note that not all Finsler functions discussed in the literature which have indefinite Finsler metrics are covered by our definition of Finsler spacetimes. As a specific example, consider the Randers type Finsler function 
\begin{equation}
F(x,y) = \sqrt{\tilde g_{ab}(x)y^ay^b} + b_a(x)y^a
\end{equation}
which is often presented as a small departure from Lorentzian metric geometry $(M,\tilde g)$ with non-vanishing one form $b$. It is clear that no smooth function $L$ on $TM\setminus\{0\}$ and no real $r\geq 2$ exist, so that $|L|^{1/r}=F$. Hence the above Finsler function does not define a Finsler spacetime in our sense. However, we cannot regard this as a problem, because the Finsler metric of the Randers Finsler function does not exist on the null cones of $\tilde g$; there, any Finsler geometric description of physics, in particular of geodesic motion, will break down. 

In contrast, the results of this paper will show that our definition of Finsler spacetimes only admits those Finsler functions for which we can control the physically relevant geometry. Before we discuss this in detail in the following sections, we wish to comment on the interpretation of fields on Finsler spacetime. 

\subsection{Interpretation}\label{sec:interpretation}
\begin{figure}[h]
\includegraphics[width=3in]{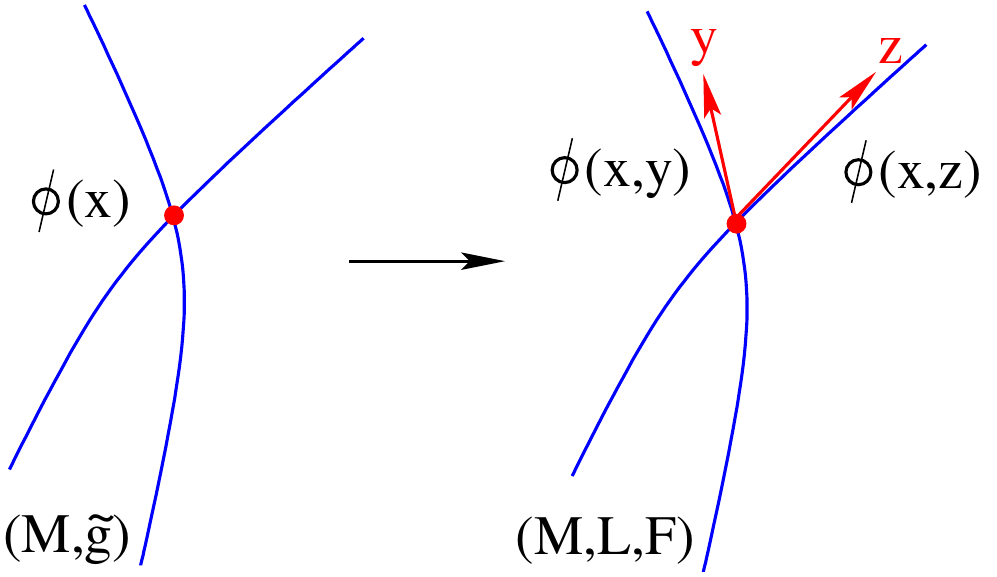}
\caption{\textit{Interpretation of fields in Lorentzian metric (dependence on position) and Finsler spacetimes (dependence on position and four-velocity).}\label{fig:0}}
\end{figure}

From the definition of Finsler spacetime we immediately realize that the metric $g^L_{ab}$ that appears as a generalization of the Lorentzian metric depends on all tangent bundle coordinates $(x,y)$, not only on the coordinates  of the manifold. Thus, in order to obtain consistent equations for physical fields $\phi$ on Finsler spacetime, it is necessary that these fields also depend on all tangent bundle coordinates. We interpret $\phi(x,y)$ as the field $\phi$ measured by an observer at the point $x\in M$ with four-velocity $y\in T_xM$, see figure~\ref{fig:0}. How fields $\phi(x,y)$ are constructed as lifts from standard fields over $M$ is discussed in more detail in section~\ref{sec:actions}.

On Lorentzian metric manifolds the observer's four velocity is responsible for the time--space split seen by the observer. With respect to this split the matter tensor field components are interpreted; in electrodynamics, for instance, the components of the field strength are interpreted as the electric and magnetic fields. 

On Finsler spacetimes the role played by the observer's four-velocity is more complicated: though it may still be responsible for a
time--space split (using orthogonality with respect to~$g^L_{ab}$), there is an additional dependence in the tensor field components. This
implies a deformation of the definitions of fields for differently moving observers at the same spacetime point.

\section{Geometry of Finsler spacetimes}\label{sec:geometry}
In order to formulate physical theories on Finsler spacetimes $(M,L,F)$ we need geometric objects such as connection, covariant derivative, curvature and tensor fields; in this section we will briefly review the standard construction of these in Finsler geometry~\cite{Bucataru, Chern}. These objects are immediately available on Finsler spacetimes wherever the Finsler metric exists and is non-degenerate; according to our definition of the previous section this is the case on $TM\setminus (A\cup \{L=0\})$. The most important new result that we will prove here shows that all standard constructions can in fact be extended to the larger domain $TM\setminus A$,
which is a consequence of the existence of the smooth function $L$ with $F^r=|L|$. In particular, this implies that all geometric objects are well-defined where $F=0$, which will allow us to discuss null geodesics and light propagation.

The Finsler function $F$ is a function on the tangent bundle $TM$, but is supposed to describe the geometry of the manifold $M$. We now wish to define distinguished tensor fields over $TM$, so-called d-tensors, that transform precisely like tensor fields over $M$ under induced coordinate transformations
\begin{equation}\label{eqn:ind}
\tilde x^a (x,y) = \tilde x^a(x)\,,\quad \tilde y^a(x,y)  = \frac{\partial \tilde x^a}{\partial x^b} y^b\,.
\end{equation}
A d-tensor $T$ takes the general form
\begin{equation}
T =  T^{a_1\dots b_1\dots}{}_{c_1\dots d_1\dots}(x,y) \,\delta_{a_1}\otimes\dots \otimes \bar\partial_{b_1}\otimes \dots \otimes dx^{c_1}\otimes \dots\otimes \delta y^{d_1}\otimes \dots
\end{equation}
in terms of the Berwald basis of $TTM$ and the corresponding dual basis of $T^*TM$,
\begin{equation}\label{eqn:Berwald}
\left\{\delta_a = \partial_a -N^b{}_a\bar \partial_b\,,\; \bar\partial_a\right\} ,\qquad
\left\{dx^a\,,\;\delta y^a = dy^a + N^a{}_b dx^b\right\}.
\end{equation}
These bases split $TTM = \langle \delta_a \rangle\oplus \langle \bar\partial_a\rangle$ and $T^*TM= \langle dx^a \rangle\oplus \langle \delta y^a\rangle$ into horizontal and vertical parts, respectively. In the definition of the Berwald bases appear the coefficients $N^a{}_b(x,y)$ of some non-linear connection. Under induced coordinate transformations~(\ref{eqn:ind}) these change as
\begin{equation}\label{eqn:Ntrans}
\tilde N^a{}_b = \frac{\partial \tilde x^a}{\partial x^p} \frac{\partial x^q}{\partial \tilde x^b} N^p{}_q + \frac{\partial\tilde x^a}{\partial x^p} \frac{\partial y^p}{\partial\tilde x^b}\,.
\end{equation}
This guarantees that the components of a d-tensor $T$ transform as required:
\begin{equation}
\tilde T^{a_1\dots b_1\dots}{}_{c_1\dots d_1\dots} = \frac{\partial\tilde x^{a_1}}{\partial x^{p_1}}\dots \frac{\partial\tilde x^{b_1}}{\partial x^{q_1}}\dots\frac{\partial x^{r_1}}{\partial\tilde x^{c_1}}\dots\frac{\partial x^{q_1}}{\partial\tilde x^{d_1}}\dots T^{p_1\dots q_1\dots}{}_{r_1\dots s_1\dots}\,.
\end{equation}
An important example for a d-tensor is the curvature tensor associated to a nonlinear connection; its components $R^c{}_{ab}(x,y)$ are defined by
\begin{equation}\label{eqn:curv}
R^c{}_{ab} =  - 2 \delta_{[a} N^c{}_{b]} =  - \partial_a N^c{}_b + \partial_b N^c{}_a + N^p{}_a \bar \partial_p N^c{}_b - N^p{}_b \bar \partial_p N^c{}_a\,.
\end{equation}

Wherever the Finsler metric of our Finsler spacetime $(M,F)$ is non-degenerate, we may now apply a standard construction to specify a unique non-linear connection in terms of $F$. This connection is called the Cartan non-linear connection and has coefficients
\begin{equation}\label{eqn:Ndef}
N^a{}_b(x,y) = \frac{1}{2}\bar\partial_b \Big(\Gamma^a{}_{pq}(x,y)y^py^q\Big)
\end{equation}
where $\Gamma^a{}_{pq}(x,y)$ are the Christoffel symbols calculated from the components $g_{ab}(x,y)$ of the Finsler metric, using $x$-derivatives as usual. The Cartan non-linear connection is the unique connection for which the Finsler metric is covariantly constant in the sense
\begin{equation}
y^p\delta_pg_{ab} - N^p{}_a g_{pb} - N^p{}_b g_{ap} =0
\end{equation}
and for which the Cartan two-form $\omega = \frac{1}{2}d\big(\bar\partial_a F^2 dx^a\big)$ vanishes on horizontal vectors, see theorem~5.5.6 in~\cite{Bucataru}. This can be compared to the construction of the Levi-Civita connection on metric spacetimes which is the unique metric compatible and torsion-free connection.

The Cartan non-linear connection can now be applied to construct a linear covariant derivative\footnote{Without entering the technical detail, we mention that this covariant derivative is the horizontal derivative via the Cartan linear connection on the tangent bundle.} that maps d-tensors with components $T^{a\dots}{}_{b\dots}$ to d-tensors with components
\begin{equation}
\nabla_c T^{a\dots}{}_{b\dots} = \delta_c T^{a\dots}{}_{b\dots} + \Gamma^{\delta\,a}{}_{pc} T^{p\dots}{}_{b\dots} + {\dots} - \Gamma^{\delta\,p}{}_{bc} T^{a\dots}{}_{p\dots} -\dots
\end{equation}
where the connection coefficients $\Gamma^{\delta\,a}{}_{bc}(x,y)$ are defined similarly as Christoffel symbols for the Finsler metric $g_{ab}$, but using horizontal derivatives:
\begin{equation}\label{eqn:Gdel}
\Gamma^{\delta\,a}{}_{bc} = \frac{1}{2} g^{ap}\Big(\delta_bg_{pc}+\delta_cg_{pb}-\delta_pg_{bc}\Big) .
\end{equation}
Under induced coordinate transformations~(\ref{eqn:ind}), these connection coefficients change in precisely the same way as the usual
metric Christoffel symbols do. The coefficients $\Gamma^{\delta\,a}{}_{bc}$ can also be used to write the curvature~(\ref{eqn:curv}) of the
Cartan non-linear connection in the more familiar form
\begin{equation}
R^c{}_{ab} = - y^d \left(\delta_a \Gamma^\delta{}^c{}_{db} - \delta_b \Gamma^\delta{}^c{}_{da} + \Gamma^\delta{}^c{}_{pa}\Gamma^\delta{}^p{}_{db} - \Gamma^\delta{}^c{}_{pb}\Gamma^\delta{}^p{}_{da}\right).
\end{equation}

After this review of standard constructions in Finsler geometry we will now proceed to show our new results. First we will establish a link between the $\Gamma^{\delta\,a}{}_{bc}$, the fibre derivatives of the Cartan non-linear connection and the totally symmetric Cartan d-tensor
\begin{equation}
C_{abc}(x,y) = \frac{1}{2} \bar\partial_a g_{bc}(x,y)\,.
\end{equation}
The following theorem will be useful for later calculations, especially for integrations by parts that occur in variations of field theory actions on Finsler spacetime.

\vspace{6pt}\noindent\textbf{Theorem 1.} 
\textit{Wherever the Finsler metric $g$ of a Finsler spacetime $(M,L,F)$ is non-degenerate, 
\begin{equation}
S^a{}_{bc} = \Gamma^{\delta\,a}{}_{bc}-\bar\partial_c N^a{}_b
\end{equation}
defines a d-tensor field. The components $S^a{}_{bc}$ can be written as $S^a{}_{bc} = - y^p\nabla_p C^a{}_{bc}$ (where the index on the Cartan tensor is raised with the inverse Finsler metric).}

\vspace{3pt}\noindent\textit{Proof.} 
The coefficients $\Gamma^{\delta\,a}{}_{bc}$ change under induced coordinate transformations~(\ref{eqn:ind}) as do the standard Christoffel symbols. The same is true for the $\bar\partial_cN^a{}_b$ as follows from~(\ref{eqn:Ntrans}). Hence their difference defines the components of a d-tensor field. In order to prove the stated relation of $S^a{}_{bc}$ to the Cartan tensor, we rewrite equation~(\ref{eqn:Ndef}) in the form
\begin{equation}
N^a{}_b = \Gamma^a{}_{bp}y^p - C^a{}_{bt}\Gamma^t{}_{pq}y^py^q\,.
\end{equation} 
Using this, the claim follows from a lengthy expansion of both expressions stated for $S^a{}_{bc}$. $\square$

\vspace{6pt}\noindent Both the Cartan tensor $C_{abc}$ and the d-tensor $S^a{}_{bc}$ vanish for a metric induced Finsler spacetime; so they measure certain aspects of the departure from metricity.

We will now prove one of the key results of this article. The geometric objects on a Finsler spacetime $(M,L,F)$ discussed above are only defined where the Finsler metric exists and is non-degenerate, i.e., on $TM\setminus (A\cup \{L=0\})$. We will prove that these geometric objects can be defined on the larger domain $TM\setminus A$; thus, in particular, they can be defined on the null structure $\{L=0\}$ of the spacetime! 

\vspace{6pt}\noindent\textbf{Theorem 2.} 
\textit{Let $(M,L,F)$ be a Finsler spacetime. Over $TM\setminus (A\cup \{L=0\})$:
\begin{enumerate}[(i)]
\item the non-linear connection coefficients $N^a{}_b[g](x,y)$  defined in (\ref{eqn:Ndef}) as functionals of the Finsler metric $g$ of $F$ can be written as functionals $\tilde N^a{}_b[g^L](x,y)$ of the Hessian $g^L$ of~$L$; the $\tilde N^a{}_b[g^L]$ are defined on $TM\setminus A$;
\item the coefficients of the linear covariant derivative fulfil $\Gamma^{\delta\,a}{}_{bc}[g] = \Gamma^{\delta\,a}{}_{bc}[g^L]$, and so they are defined on $TM\setminus A$.
\end{enumerate}}

\vspace{3pt}\noindent\textit{Proof.} 
The strategy to prove part (i) of the theorem is to define
\begin{equation}\label{eqn:obj}
\tilde \Gamma^a{}_{bc}[g^L] = \frac{1}{2(r-1)}g^{L\,ap}\Big(\partial_bg^L_{pc}+\partial_cg^L_{pb}-\frac{2}{r}\partial_pg^L_{bc}\Big)
\end{equation}
for which we can show that on $TM\setminus (A\cup \{L=0\})$:
\begin{equation}\label{eqn:pr1}
\tilde \Gamma^a{}_{bc}[g^L]y^by^c=\Gamma^a{}_{bc}[g]y^by^c\,.
\end{equation}
According to definition (\ref{eqn:Ndef}) we then obtain the desired result that the coefficients of the
Cartan non-linear connection can be expressed as functionals of $g^L$. Since the inverse of $g^L$ appears in~(\ref{eqn:obj}) the definition of the Cartan non-linear connection extends to $TM\setminus A$ where $g^L$ is non-degenerate. 

In order to validate (\ref{eqn:pr1}) we use the homogeneity of $F$ to rewrite the right hand side as 
\begin{equation}
\frac{1}{2}g^{ap}\Big(y^b\partial_b \bar\partial_p F^2 - \partial_p F^2\Big).
\end{equation} 
Now we replace $F^2=|L|^{2/r}$ and express $g^{ap}$ as in (\ref{eqn:ggL}). This yields
\begin{equation}\label{eqn:useful}
\Gamma^a{}_{bc}y^by^c = \frac{1}{2} g^{L\,ap} y^b\partial_b\bar\partial_p L -\frac{1}{2} g^{L\,ap}\partial_p L\,,
\end{equation}
which is recognized as a rewriting of the desired expression $\tilde \Gamma^a{}_{bc}[g^L]y^by^c$ by using the homogeneity properties of $L$. 

In order to prove part (ii) of the theorem, we simply need to replace all occurrences of $g_{ab}$ in the definition~(\ref{eqn:Gdel}) by
$g^L_{ab}$ using the relations in~(\ref{eqn:ggL}). Expanding all terms then yields the result. $\square$

\vspace{6pt}\noindent Since all geometric objects discussed in this section depend on the choice of non-linear connection~$N^a{}_b$ and on $\Gamma^{\delta \,a}{}_{bc}$, they are now seen to be well-defined on $TM\setminus A$ which includes the full null structure of the underlying Finsler spacetime.

When we discuss Lorentzian metric manifolds as a special case of Finsler spacetimes in section~\ref{sec:examples}, we will see that the Cartan non-linear connection reduces to the Levi-Civita connection; accordingly also the curvature is reduced to the usual Riemann curvature.

\section{Causal structure}\label{sec:causal}
In this section we will describe the causal structure of Finsler spacetimes~$(M,L,F)$. It will be shown that their definition implies a precise notion of timelike vectors that can be used to distinguish the tangents to observers' worldlines. We will prove that these timelike vectors lie in open convex cones similarly as in Lorentzian metric geometry. These cones are bounded by null vectors with $L=F=0$, but the complete null structure of a Finsler spacetime may be considerably more complicated. Moreover, the standard discussion of Finsler geodesics in terms of~$F$, which breaks down on the null structure, is generalized. We will present an improved description in terms of $L$ that is applicable almost everywhere on $TM$, especially on the null structure. We thus show that the motion of massive observers and particles and the expected motion of light are well-defined on Finsler spacetimes. That light indeed propagates on null Finsler geodesics will be confirmed in section~\ref{sec:ED} by an explicit construction of a theory of electrodynamics.

\subsection{Timelike cones}
The unit timelike condition in our definition of Finsler spacetimes intuitively provides a shell~$S_x$ of unit timelike vectors at each point $x$. The following theorem shows that this shell can be rescaled to form an open convex cone $C_x$ of timelike vectors.

\vspace{6pt}\noindent\textbf{Theorem 3.} 
\textit{Each tangent space $T_xM$ of a Finsler spacetime $(M,L,F)$ contains an open convex cone
 \begin{equation}
 C_x = \bigcup_{\lambda>0} \lambda S_x =  \bigcup_{\lambda>0} \left\{\lambda u \,|\, u\in S_x\right\}.
 \end{equation}}

\vspace{3pt}\noindent\textit{Proof.} The techniques for this proof are adapted from Beem~\cite{Beem}. To begin note that the shell of unit timelike vectors $S_x$ is a three-dimensional closed submanifold of $T_xM\cong \mathbb{R}^4$. We now proceed in three steps. First we will determine the normal curvatures of $S_x$ at some point $y_0\in S_x$. These are defined~\cite{Spivak} as $\kappa_n(z) = \sum_a \ddot \gamma^a(0)n^a$ for curves $\tau\mapsto \gamma(\tau)$ in $S_x$ with normalized tangent vectors $\sum_a \dot\gamma^a\dot\gamma^a=1$ and initial conditions $\gamma^a(0)=y^a_0$ and $\dot \gamma^a(0)=z^a$. The initial tangent $z$ is tangent to~$S_x$ in~$y_0$, i.e., it satisfies
\begin{equation}\label{eqn:yz}
0 = \bar\partial_a |L| (x,y_0) z^a = \frac{2|L(x,y_0)|}{(r-1)L(x,y_0)} g^L_{ab}(x,y_0)z^ay_0^b\,.
\end{equation}
The unit normal is given by
\begin{equation}
n^a = \frac{1}{N(x,y_0)} \bar\partial_a |L|(x,y_0) = \frac{|L(x,y_0)|g^L_{ab}(x,y_0)y_0^b}{L(x,y_0)\left(\sum_c g^L_{cp}(x,y_0)g^L_{cq}(x,y_0)y_0^py_0^q\right)^{1/2}} \,.
\end{equation}
Since the curves $\gamma(\tau)$ lie in $S_x$ where $|L|=1$, we may obtain a useful relation for $\ddot\gamma(0)$ by differentiating $|L(x,\gamma(\tau))|=1$; we find that $g^L_{ab}(x,y_0)y_0^b\ddot\gamma^a(0) = -(r-1) g^L_{ab}(x,y_0) z^az^b$. Combining these results we find the normal curvatures
\begin{equation}
\kappa_n(z) = - \frac{(r-1)|L(x,y_0)|}{N(x,y_0)L(x,y_0)} g^L_{ab}(x,y_0)z^az^b\,.
\end{equation}

Second we will show that all these normal curvatures are positive. Note that the homogeneity of $L$ implies $g^L_{ab}(x,y_0)y_0^ay_0^b=\frac{1}{2}r(r-1)L(x,y_0)$ so that $y_0$ is $g^L(x,y_0)$-timelike. Due to~(\ref{eqn:yz}) we know that $z$ and $y_0$ are $g^L(x,y_0)$-orthogonal, hence $z$ must be $g^L(x,y_0)$-spacelike, i.e., $\textrm{sign}(g^L_{ab}(x,y_0)z^az^b)=-|L(x,y_0)|/L(x,y_0)$. This immediately confirms the result $\kappa_n(z) > 0$.

Finally we will show the convexity of the set $C_x$ defined in the theorem. Positivity of the normal curvatures implies positivity of the principal curvatures of $S_x$. Now the set $\tilde C_x^1=\bigcup_{\lambda\ge 1}\lambda S_x$ is closed, connected and convex with boundary $S_x$. Because of the homogeneity of $L$, we then conclude that for all $\mu>0$ the sets $\tilde C_x^\mu=\bigcup_{\lambda\ge \mu}\lambda S_x$ are closed, connected and convex. But then 
\begin{equation}
\bigcup_{\mu>0} \tilde C_x^\mu = C_x
\end{equation}
is an open convex cone, which concludes the proof. $\square$

\vspace{6pt}\noindent So Finsler spacetimes provide a precise notion of timelike vectors: $y\in T_xM$ is called timelike if and only if $y\in C_x$. We will use these vectors in the following section to define timelike geodesics as the worldlines of observers.

Since $C_x\subset T_xM$ is an open convex cone, we may define the dual cone in $T^*_xM$ by
\begin{equation}
C^*_xM = \big\{p \in T^*_xM\, \big| \, p(y) > 0 \textrm{ for all } y\in C_x  \big\}\,.
\end{equation}
The dual cone is also open and convex, and clearly defines the set of physical momenta $p$ with positive energy with respect to an arbitrary observer, compare~\cite{Raetzel:2010je}.

Next we prove a result that characterizes in particular the boundary of the cone $C_x$ of timelike vectors. As in Lorentzian metric geometry we will find that this boundary is built from null vectors $y\in T_xM$ for which $L(x,y)=0$.

\vspace{6pt}\noindent\textbf{Theorem 4.} 
\textit{Let $(M,L,F)$ be a Finsler spacetime. Consider a connected component $T_x$ of 
\begin{equation}
\big\{y\in T_xM\,\big|\, |L(x,y)|=1\,,\;g^L_{ab}(x,y)\textrm{ is non-degenerate}\big\}
\end{equation}
and define the open set $\tilde T_x = \bigcup_{\lambda>0} \lambda T_x$. Then for $(x,y)$ in the boundary $\partial \tilde T_x$ we either have that $L(x,y)=0$ or that $g^L_{ab}(x,y)$ is degenerate.}

\vspace{3pt}\noindent\textit{Proof.} Using the homogeneity of $L$ we can write $\tilde T_x$ as
\begin{equation}
\big\{y\in T_xM\,\big|\,\exists \lambda>0 : |L(x,y)|=\lambda^r\,,\;g^L_{ab}(x,y)\textrm{ is non-degenerate}\big\}
\end{equation}
The boundary $\partial \tilde T_x$ hence consists of vectors $y\in T_xM$ with $L(x,y)=0$ or degenerate $g^L_{ab}(x,y)$. $\square$

\vspace{6pt}\noindent The proof shows in particular that the boundary of the open convex cone $C_x$ is null. This follows from the additional fact that $S_x$ is closed; then the boundary of $C_x$ must be null, and hence $g^L_{ab}$ cannot be degenerate.

\subsection{Finsler geodesics}\label{sec:ppact}
From the physical point of view the Finsler length integral  (\ref{eqn:Flength}) provides an action integral for the motion of massive observers and point particles. Their worldlines are described by the geodesics that extremize this integral. In the standard treatment the geodesic equations are the Euler--Lagrange equations
\begin{equation}
\frac{d}{d\tau}\frac{\partial F}{\partial \dot x^a}-\frac{\partial F}{\partial x^a} = 0\,.
\end{equation}
Using the relation $g_{ab}(x,y)y^ay^b=F(x,y)^2$ between the Finsler function and the Finsler metric, it is not difficult to express the geodesic equations in arclength parametrization through the coefficients $N^a{}_b$ of the Cartan non--linear connection
\begin{equation}\label{eqn:Fgeo}
\dot x^b\nabla_b \dot x^a = \ddot x^a + N^a{}_b(x,\dot x) \dot x^b = 0\,.
\end{equation}
These equations are not defined where $F$ is not differentiable, hence they cannot be applied to null motion. This is consistent with the fact that the action (\ref{eqn:Flength}) vanishes for $F=0$ and so its variation cannot imply any equation.

We will now show that the Finsler function $L$ can be used to describe the motion of all point particles, including massive observers and particles as well as the expected effective motion of light. Consider the following action integral for a curve $\tau \mapsto x(\tau)$ in $M$,
\begin{equation}\label{eqn:Lact}
S[x] = \int d\tau \, \Big(L(x(\tau),\dot x(\tau)) + \lambda(\tau) \big[L(x(\tau),\dot x(\tau)) - \kappa\big]\Big)
\end{equation}
where a normalization constant $\kappa=0,\pm 1$ appears. This action formulation does not involve mass explicitly, which is consistent with the weak equivalence principle. 

In the Lorentzian metric case $L(x,\dot x)=\tilde g_{ab}(x)\dot x^a\dot x^b$, one recognizes the standard quadratic point particle action; the Lagrange multiplier $\lambda$ then controls whether the resulting geodesics are massive with unit timelike tangent vectors $\tilde g_{ab}\dot x^a\dot x^b=\kappa=-1$, spacelike with $\kappa=+1$, or massless with null tangent vectors $\kappa=0$. 

Also for general $L$, the constraint from the variation of the action with respect to the Lagrange multiplier $\lambda$ ensures arclength parametrization
\begin{equation}
L(x(\tau),\dot x(\tau)) = \kappa
\end{equation}
which is equivalent to $F=|\kappa|$ along the curve. Massless motion will again be described by $\kappa=0$, while massive motion will be described by the timelike vectors $\dot x \in C_x$ for which $|\kappa|=1$. The requirement that the variation of the action with respect to the curve $x(\tau)$ should vanish is the Euler-Lagrange equations
\begin{equation}
\frac{d}{d\tau}\Big((1+\lambda)\frac{\partial L}{\partial \dot x^a}\Big)-(1+\lambda)\frac{\partial L}{\partial x^a} = 0\,.
\end{equation}
With the help of (\ref{eqn:gLL}) and Theorem~2 we rewrite these equations in the equivalent form
\begin{equation}\label{eqn:EL}
\dot x^b\nabla_b \dot x^a = -\frac{\dot \lambda}{(r-1)(1+\lambda)} \dot x^a
\end{equation}
In case $\kappa = \pm 1$, the parametrization condition implies $g^L_{ab}(x,\dot x)\dot x^a \dot x^c\nabla_c\dot x^b=0$, so we conclude
from equation (\ref{eqn:EL}) that $\dot \lambda = 0$ and obtain the Finsler geodesic equation (\ref{eqn:Fgeo}). For null curves $\kappa=0$ we can simply reparametrize $\tau\mapsto \sigma(\tau)$ without changing $L(x,x')=0$ to obtain the Finsler geodesic equation (\ref{eqn:Fgeo}) in terms of derivatives with respect to the parameter $\sigma$.

It is important to note that the derivation of Finsler geodesics from the action (\ref{eqn:Lact}) involving $L$ also results in the Cartan non-linear connection expressed through derivatives of $L$, see Theorem~2 in section~\ref{sec:geometry}. Hence we can formulate Finsler geodesics with this action on $TM\setminus A$ which includes the full null structure $\{L=0\}$.

We conclude that the technology available on Finsler spacetimes as we have defined them in section~\ref{sec:spacetime} enables us to describe massive observers and point particles as Finsler geodesics $\tau\mapsto x(\tau)$ in $M$ with timelike tangent vectors $\dot x(\tau)\in C_{x(\tau)}$. Moreover, we can describe the expected effective motion of light or massless particles by well-defined null Finsler geodesics with $L(x(\tau),\dot x(\tau))=0$. This expectation will be justified field-theoretically in section~\ref{sec:ED}.

\section{Illustrative examples}\label{sec:examples}
After these technical preparations we are now in the position to discuss in detail two simple examples of Finsler spacetimes $(M,L,F)$. These illustrate the strength of our definition and the general theorems derived above. First we will show that Lorentzian metric spacetimes are a special case of Finsler spacetimes. In particular we will exhibit how connection and curvature, and the causal structure of a Lorentzian metric fit into the more general scheme discussed above. The second example shows a more complicated causal structure with two different lightcones at each point. This Finsler spacetime goes beyond metric manifolds, but nevertheless has well-defined timelike cones and allows a full description of observers and null motion.

\subsection{Lorentzian metric spacetimes}
Lorentzian manifolds $(M,\tilde g)$ with metric $\tilde g$ of signature $(-,+,+,+)$ are a special type of Finsler spacetimes $(M,L,F)$. They are described by  the metric-induced function
\begin{equation}
L(x,y) = \tilde g_{ab}(x)y^ay^b
\end{equation}
which is homogeneous of degree $r=2$. Recalling the definition, $L(x,y)$ leads to the Finsler function $F(x,y) = |\tilde
g_{ab}(x)y^ay^b|^{1/2}$ that is easily recognized as the integrand of the Lorentzian length as described in the motivation of this article. 

\begin{figure}[h]
\includegraphics[width=2in]{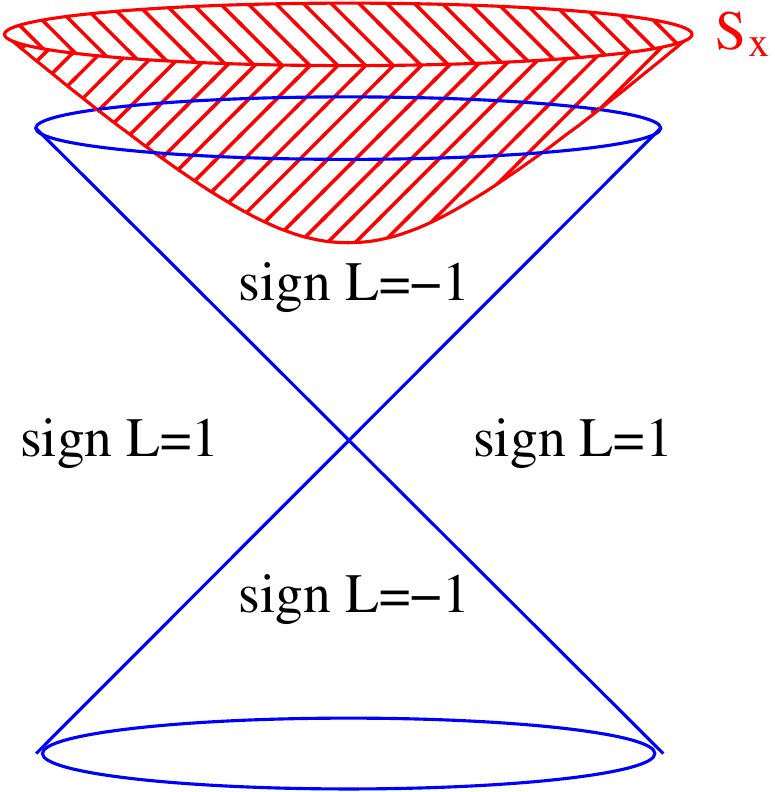}\hspace{0.5in}\includegraphics[width=2.6in]{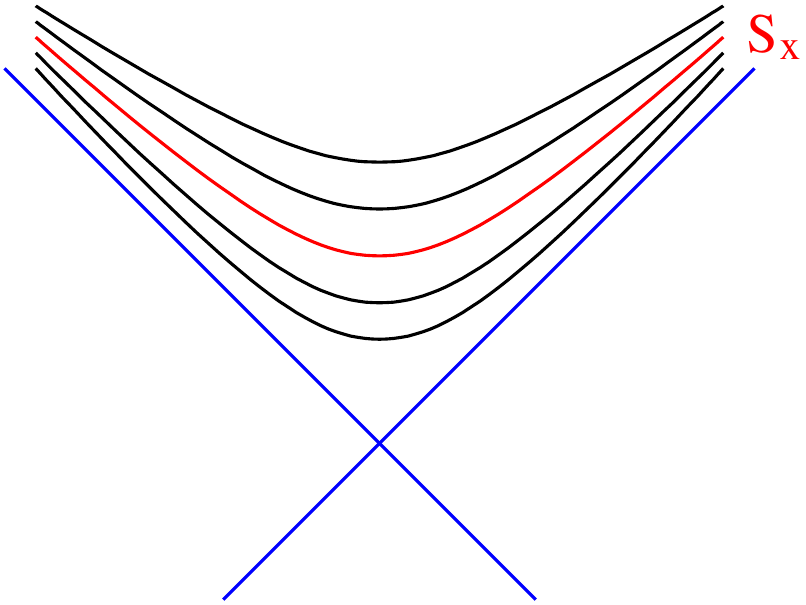}
\caption{\textit{\\ (a) Left: null structure and shell of unit timelike vectors on Lorentzian metric spacetimes.\\ 
(b) Right: rescaling the unit shell to form the cone of timelike vectors.}\label{fig:1}}
\end{figure}

Clearly $L$ is smooth on $TM$ and obeys the reversibility property. The metric $g^L_{ab}(x,y)=\tilde g_{ab}(x)$, and hence is non-degenerate on $TM$; so the measure zero set $A=\emptyset$. The signature of $g^L$ is globally $(-,+,+,+)$, so the unit timelike condition tells us to consider the set 
\begin{equation}
\Omega_x=\Big\{y\in T_xM\,\Big|\,\epsilon(x,y)=\frac{|L(x,y)|}{L(x,y)}=-1 \Big\}\,.
\end{equation}
This set has precisely two connected components, both of which are closed. We may call one of these $S_x$, as displayed in 
figure~\ref{fig:1}(a). From Theorem~3 in section~\ref{sec:causal} we learn that the shell of unit timelike vectors $S_x$ can be rescaled to
form an open convex cone~$C_x$ that contains all the usual timelike vectors of $\tilde g$ at a point $x\in M$, see figure~\ref{fig:1}(b). 

The Finsler function $F$ is non-differentiable on the null structure $\{L=0\}\subset TM$; there the Finsler metric is not defined. We have the results $g_{ab}(x,y)=-\tilde g_{ab}(x)$ on the $\tilde g$-timelike vectors and $g_{ab}(x,y)=+\tilde g_{ab}(x)$ on the $\tilde g$-spacelike vectors, hence the Finsler metric changes its signature.

The geometric constructions of section~\ref{sec:geometry} take the familiar form for Lorentzian manifolds. The Cartan tensor $C_{abc}$ and the tensor $S^a{}_{bc}$ that measure the departure from metricity vanish, because the Finsler metric does not depend on the fibre coordinates. Moreover, the coefficients of the linear covariant derivative simply become the Christoffel symbols of the metric $\tilde g$, i.e., $\Gamma^{\delta\,a}{}_{bc}(x,y) = \Gamma^a{}_{bc}(x)$. The non-linear connection reduces to a linear connection with coefficients
\begin{equation}
N^a{}_b(x,y) = \Gamma^a{}_{bc}(x)y^c\,, 
\end{equation}
and according to~(\ref{eqn:curv}) its curvature is given by the Riemann tensor of $\tilde g$ as $R^c{}_{ab}(x,y) = - y^d R^a{}_{dbc}(x)$. 

For Finsler spacetimes induced by Lorentzian metrics it is easy to see that connection and curvature are expressible in terms of $g^L_{ab}(x,y)=\tilde g_{ab}(x)$ and hence defined everywhere on $TM$, not only where the Finsler metric is defined. This is a very special case of Theorem~2 of section~\ref{sec:geometry}.

\subsection{Simple bimetric Finsler structure}
A simple example of a Finsler spacetime $(M,L,F)$ that goes beyond Lorentzian metric manifolds can be defined through two Lorentzian metrics $h$ and $k$ of signature $(-,+,+,+)$ for which the cone of $h$-timelike vectors is contained and centred in the cone of $k$-timelike vectors. As mentioned before, such Finsler spacetimes are relevant as covariant descriptions for certain aspects of crystal optics. It is worth noting that it was thought impossible to realize two signal cones consistently in Finsler geometry~\cite{Skakala:2010hw}, but we will see that this is not a problem at all. Our example is based on the function
\begin{equation}
L=h_{ab}(x)y^ay^b\,k_{cd}(x)y^cy^d\,.
\end{equation}
It is clear that $L$ is homogeneous of degree $r=4$, smooth on $TM$ and obeys the reversibility condition. The corresponding Finsler
function is defined as $F(x,y)=|h_{ab}(x)y^ay^bk_{cd}(x)y^cy^d|^{1/4}$. 

\begin{figure}[h]
\includegraphics[width=2.5in]{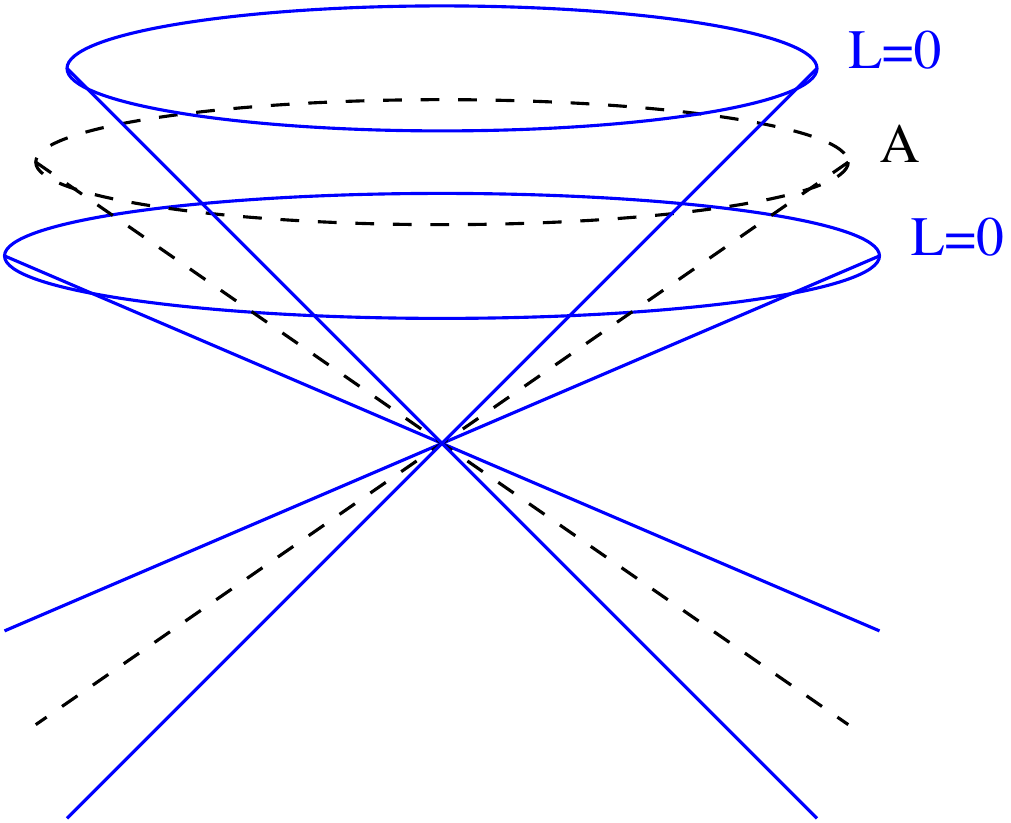}
\caption{\textit{Null structure of the bimetric Finsler spacetime (solid) and degeneracy set $A$ of $g^L$ (dashed).}\label{fig:2}}
\end{figure}

The null structure $\{L=0\}$ is the union of the null cones of the metrics $h$ and $k$, and the metric $g^L_{ab}(x,y)$ turns out to be degenerate on a measure zero subset $A\neq \emptyset$ that forms an additional structure between the null surfaces, as displayed in figure~\ref{fig:2}. 

\begin{figure}
\includegraphics[width=3in]{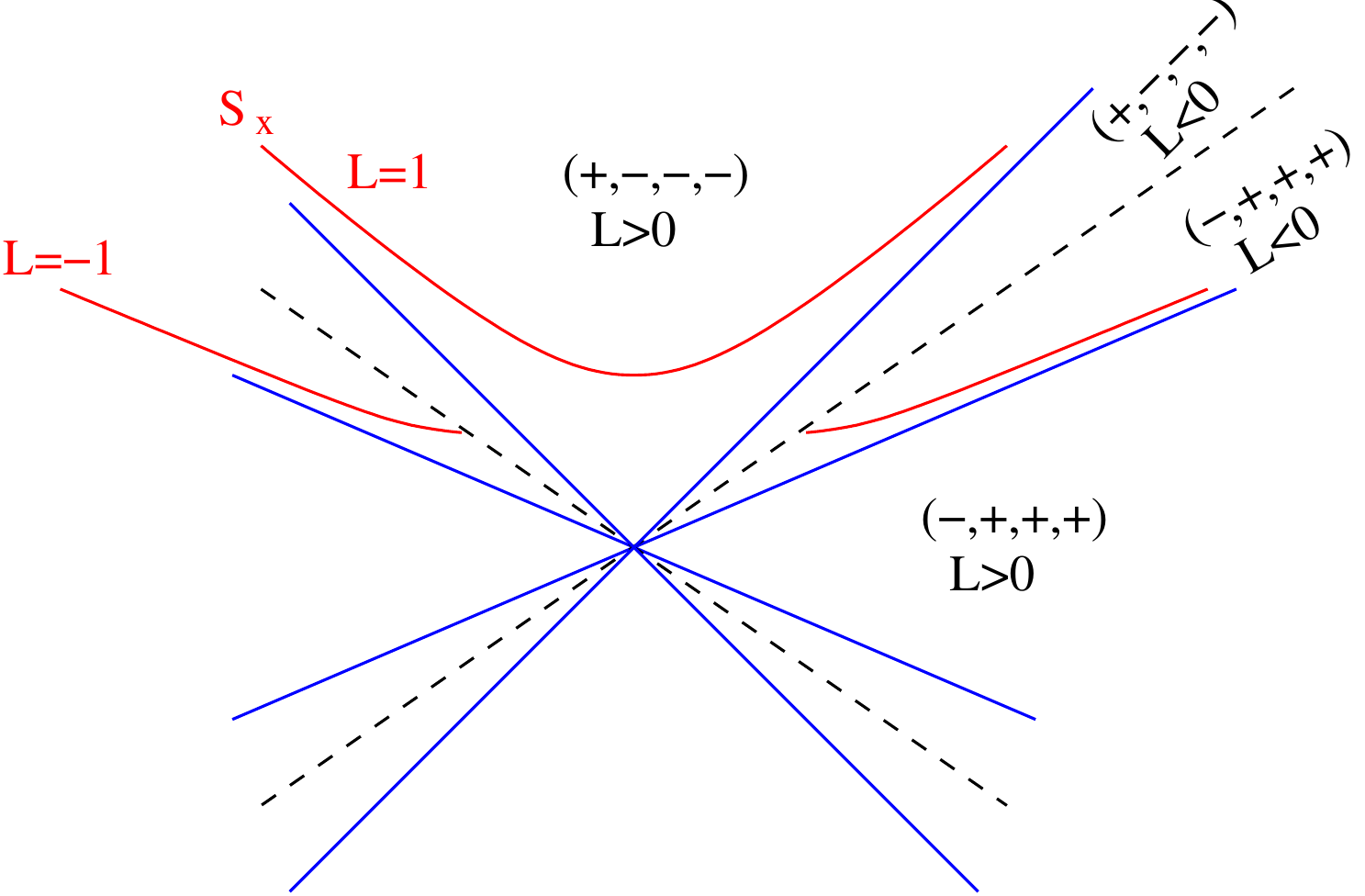}\hspace{0.5in}\includegraphics[width=2.5in]{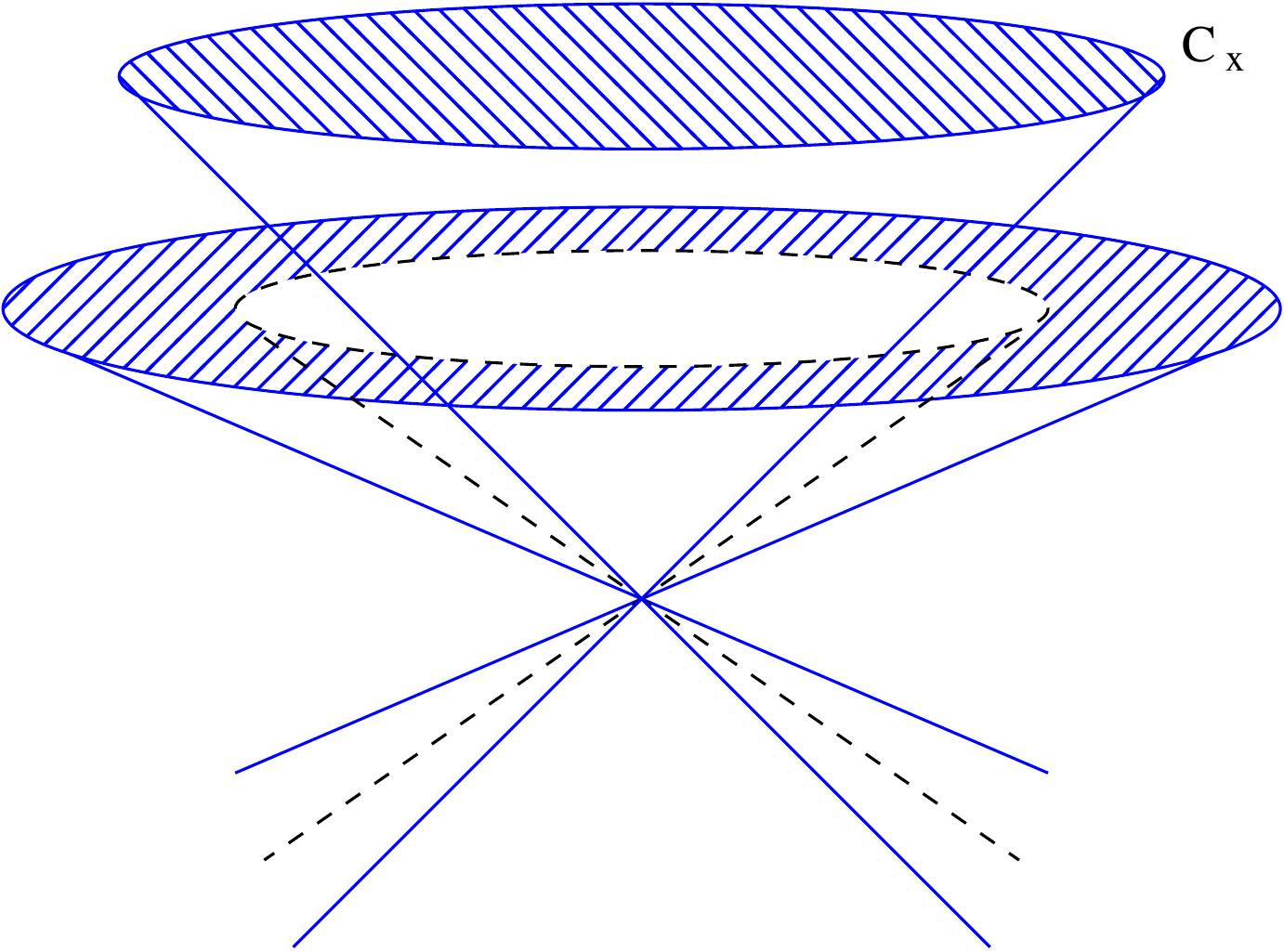}
\caption{\textit{\\(a) Left: cut through two connected components of $\Omega_x$; the inner set $S_x$ is closed, the outer set is not.\\ 
(b) Right: rescaling $S_x$ leads to a convex cone $C_x$; rescaling the outer set does not.}\label{fig:3}}
\end{figure}

Across $A$, the metric $g^L$ changes its signature from $(+,-,-,-)$ to $(-,+,+,+)$. In order to analyse the unit timelike condition, we need to
compare the signature of $g^L$ with the sign of $L$. One finds four connected components of the set $\Omega_x$. Two of these are closed, two
are not; one of each is displayed in figure~\ref{fig:3}(a). Choosing one of the closed components to be the set $S_x$ we can rescale it to form
the complete convex cone $C_x$ of timelike vectors at $x\in M$ according to Theorem~3 of section~\ref{sec:causal}. The non-closed components
will not give rise to a convex cone when rescaled in the same way, as can be seen in figure~\ref{fig:3}(b).

As in the Lorentzian metric case the Finsler function $F$ of this Finsler spacetime is not differentiable where $L=0$.  There the Finsler metric is not defined; it changes its signature across the null structure and is degenerate on~$A$.  It  is necessary to apply Theorem~2 of section~\ref{sec:geometry} to realize that the Cartan non-linear connection and its curvature are well-defined on $TM\setminus A$, in particular on the set $\{L=0\}$. Hence this bimetric Finsler spacetime has a well defined causal structure which is more general than that of Lorentzian metric manifolds but admits all the necessary properties to be applicable in physics. So Finsler spacetimes are indeed  nice generalizations of Lorentzian metric manifolds. In the next section we will demonstrate that it is possible to formulate physical field theories on these generalized backgrounds.

\section{Field theory actions on Finsler spacetimes}\label{sec:actions}
Classical physics is described by field theories and effective massive and massless point particles on Lorentzian metric spacetimes. Since Finsler spacetimes present a natural generalization of metric spacetimes, we should not only be able to formulate point particle actions, as we already did in section~\ref{sec:ppact}, but also generalized field theories. In this section we will develop the technology needed to write down well-defined field theory action integrals, and to derive the corresponding equations of motion. We will also argue why certain past attempts to formulate such actions on Finsler spaces are technically incorrect.

An extended field $\phi$ on a Finsler spacetime is a tensor field with definite homogeneity on the tangent bundle. Hence the components of $\phi$ are functions of all tangent bundle coordinates, $\phi^{A\dots}{}_{B\dots}(x,y)$, which are measured by observers moving through a spacetime point $x$ with worldline tangent $y$, as argued in section~\ref{sec:interpretation}. Naively, an action $S[\phi]$ for $\phi$ would be an integral over some Lagrangian density $\tilde{\mathcal{L}}[\phi](x,y)$ on~$TM$ with variation
\begin{equation}\label{eqn:vari}
\delta S[\phi] = \int_{TM} d^4xd^4y\, \delta \tilde{\mathcal{L}}[\phi](x,y) = \int_{TM} d^4xd^4y\,\frac{\delta S[\phi]}{\delta \phi^{A\dots}{}_{B\dots}(x,y)} \delta \phi^{A\dots}{}_{B\dots}(x,y)\,.
\end{equation}
However, if the integrand is of definite homogeneity~$n$, the variation $\delta S[\phi]$ always diverges. Therefore one cannot require $\delta S[\phi]=0$ in order to read off equations of motion; this issue was not appreciated in~\cite{Voicu:2009wi,Voicu:2010kw}.

The divergence of the variation $\delta S[\phi]$ becomes clear by using special coordinates $(\hat x^a, u^\alpha,R)$ on the domain $\{L\neq 0\}\subset TM$, which will be constructed in detail below, to write~(\ref{eqn:vari}) as
\begin{eqnarray}
\int_{\{L=0\}} d^4x d^4y \, \delta \tilde{\mathcal{L}}[\phi](x,y)
+\Big( \int_0^\infty dR\, R^{n+3}\Big) \int d^4\hat x d^3u \,J(\hat x,u,1)\, \delta\tilde{\mathcal{L}}[\phi](\hat x,u,1)
\end{eqnarray}
where $J$ is the determinant of the Jacobian of the coordinate change. The integral over $R$ absorbs the $n$-homogeneity of the integrand and clearly diverges. Note that this problem cannot be cured by considering compactly supported $\delta \phi^{A\dots}{}_{B\dots}(x,y)$; these simply do not exist because homogeneity always leads to non-compact support along the fibre directions.

We will now present the technology to formulate well-defined action integrals for fields on Finsler spacetimes. As we have seen, these actions cannot be integrals over the whole tangent bundle because of homogeneity. The central idea is to divide out the homogeneity by restricting the integration to the subbundle of  constant $|L|$ in $TM$. Variations of the so constructed actions will lead to equations of motion on the same subbundle. The homogeneity of all involved fields then guarantees that the equations can be extended back to the whole tangent bundle.

More precisely, we formulate actions  for homogeneous fields $\phi$ on Finsler spacetimes as integrals over the subbundle $\Sigma\subset (TM\setminus A)\subset TM$ that is defined as
\begin{equation}
\Sigma  = \bigcup_{x\in M}\big\{y\in T_xM\,\big|\, |L(x,y)|=1\,,\;g^L_{ab}(x,y)\textrm{ is non-degenerate}\big\}\,.
\end{equation}
For explicit calculations it will be convenient to introduce coordinates on $\Sigma$. The charts we consider live on the smooth choice of a connected component of $\Sigma_{|x}$ over an open set $x\in U\subset M$. Similarly as in Asanov~\cite{Asanov} we start from induced coordinates $(Z^A)=(x^a,y^a)$ and define new coordinates $(\hat Z^A)=(\hat x^a,u^\alpha,R)$ on $TM$ that satisfy
\begin{equation}
\hat x^a(x,y) = x^a\,,\qquad
R(x,y) = |L(x,y)|^{1/r}\,,\qquad
u^\alpha(x,y) \textrm{ homogeneous of degree zero in }y\,.
\end{equation}
It will not be necessary to specify the coordinates $u^\alpha$ more explicitly. We simply note that they can be constructed from the zero-homogeneous functions $y^a/|L(x,y)|^{1/ r}$ that provide an embedding~$\Sigma\hookrightarrow TM$.

We will now deduce a number of properties of the new coordinates. For an $n$-homogeneous function $h$ on $TM$ with $h(x,\lambda y)=\lambda^nh(x,y)$ for positive $\lambda$, we find that the homogeneity with respect to the fibre coordinates is translated into homogeneity with respect to $R$. Indeed,
\begin{equation}
\lambda^nh(\hat x^a(x),\hat u^\alpha(x,y),R(x,y))=h(\hat x^a(x),\hat u^\alpha(x,y),R(x,\lambda y))\,;
\end{equation} 
then differentiation with respect to $\lambda$ at $\lambda=1$ shows that $h$ is $n$-homogeneous in $R$ by Euler's theorem. In particular this implies that $y^a$ is one-homogeneous, $R\,\partial_Ry^a = y^a$. By direct calculation we thus obtain from the coordinate transformation the basis change matrices on $TTM$:
\begin{equation}
\frac{\partial \hat Z^A}{\partial Z^B} = \left[\begin{array}{c|c} \delta^a_b&0\\ \hline \begin{array}{c} \partial_b u^\alpha \\ \partial_b |L|^{1/r} \end{array}& \begin{array}{c} \bar\partial_b u^\alpha \\ \bar\partial_b |L|^{1/r} \end{array}\end{array}\right]\,,\qquad
\frac{\partial Z^A}{\partial \hat Z^B} = \left[\begin{array}{c|c} \delta^a_b&\begin{array}{cc}0\quad& \;0 \end{array}\\ \hline \hat\partial_by^a& \begin{array}{cc} \partial_{u^\beta} y^a & \frac{y^a}{R} \end{array} \end{array}\right].
\end{equation}
These matrices are inverses of one another, so we can read off many relations that will become important when we will deduce a canonical volume form on $\Sigma$, and perform variations of the field theory actions:
\begin{equation}\label{eqn:inv1}
\frac{\partial \hat Z^A}{\partial Z^C}\frac{\partial Z^C}{\partial\hat Z^B} = 
\left[\begin{array}{c|c}\delta^a_b & 0 \\ \hline \begin{array}{c} \partial_b u^\alpha+\bar\partial_c u^\alpha \hat\partial_b y^c \\ \partial_b |L|^{1/r} + \bar\partial_c|L|^{1/r} \hat\partial_b y^c \end{array}& \begin{array}{cc} \bar\partial_c u^\alpha \partial_{u^\beta} y^c & \bar\partial_c  u^\alpha \frac{y^c}{R} \\ \bar\partial_c|L|^{1/r} \partial_{u^\beta} y^c & 1 \end{array}\end{array}\right] = 
\left[\begin{array}{c|c}\delta^a_b & 0 \\ \hline \begin{array}{c}0 \\ 0 \end{array}& \begin{array}{cc} \delta^\alpha_\beta & 0 \\ 0 & 1 \end{array}\end{array}\right] ,
\end{equation}

\begin{equation}\label{eqn:inv2}
\frac{\partial Z^A}{\partial \hat Z^C}\frac{\partial \hat Z^C}{\partial Z^B} = 
\left[\begin{array}{c|c}\delta^a_b & 0 \\ \hline  \hat \partial_b y^a + \partial_b u^\gamma\partial_{u^\gamma} y^a + \frac{y^a}{R} \partial_b |L|^{1/r} & \partial_{u^\gamma} y^a \bar\partial_b u^\gamma + \frac{y^a}{R} \bar\partial_b|L|^{1/r} \end{array}\right] = 
\left[\begin{array}{c|c}\delta^a_b & 0 \\ \hline 0& \delta^a_b \end{array}\right] .
\end{equation}

In physical field theory actions one usually writes the Lagrangian density $\tilde{\mathcal{L}}[\phi](x,y)$ as a volume form multiplied
with a scalar Lagrangian $\mathcal{L}[\phi](x,y)$. We construct the volume form on our integration domain $\Sigma\subset TM\setminus A$ as
the pullback of a very simple volume form on $TM\setminus A$ which is determined by the Sasaki type $r$-homogeneous metric
\begin{eqnarray}\label{eqn:GTM}
G & =&  |L|^{2/r} g^L_{ab} dx^a\otimes dx^b +g^L_{ab} \delta y^a\otimes \delta y^b\nonumber\\
& =& R^{2} g^L_{ab} d\hat x^a\otimes d\hat x^b + h_{\alpha\beta} \delta u^\alpha \otimes \delta u^\beta + \frac{r(r-1)L}{2\,R^2} dR\otimes dR
\end{eqnarray}
where $h_{\alpha\beta} = \partial_{u^\alpha} y^a \partial_{u^\beta} y^b g^L_{ab}$ and $\delta u^\alpha = du^\alpha + (\bar\partial_b u^\alpha N^b{}_a -  \partial_a u^\alpha)d\hat x^a$. We remark that one can determine the signature of $h$, given the signature of $g^L$ and the sign of $L$.\footnote{For instance, on the cone $C_x$ of timelike vectors in the bimetric example of section~\ref{sec:examples} we have positive $L>0$ and the signature $(+,-,-,-)$ of $g^L$. 
There it follows that $h$ has definite signature $(-,-,-)$.} The pullback of the metric~$G$ to $\Sigma$ is simply obtained by setting $R=1$; in local coordinates this implies the following volume form on $\Sigma$:
\begin{equation}
\left(\sqrt{|\textrm{det }g^L_{ab}\, \textrm{det }h_{\alpha\beta}|}\right)_{|\Sigma} d\hat x^0\wedge d\hat x^1\wedge d\hat x^2 \wedge d\hat x^3 \wedge du^1\wedge du^2 \wedge du^3\,.
\end{equation}
The restriction of a function $f$ on $TM$ to $\Sigma$ is always obtained by setting $R=1$, i.e.,
\begin{equation}
f_{|\Sigma}(\hat x,u) = f(\hat x,u,1)\,.
\end{equation}
Combining these arguments, we conclude that field theory actions for  physical fields $\phi$ on a Finsler spacetime take the general form
\begin{equation}\label{eqn:fact}
S[\phi] = \int_\Sigma d^4\hat x d^3u\,\left(\sqrt{g^L\, h}\;\mathcal{L}[\phi]\right)_{|\Sigma}
\end{equation}
where we use the shorthand notation $g^L=|\textrm{det }g^L_{ab}|$ and $h=|\textrm{det }h_{\alpha\beta}|$.

Physical scalar Lagrangians $\mathcal{L}[\phi]$ depend locally on the field $\phi$ and its derivatives up to some finite order. To derive equations of motion by variation of the field theory action~(\ref{eqn:fact}) we hence need to know how to perform integrations by parts. Using the relations in~(\ref{eqn:inv1}) and~(\ref{eqn:inv2}) one can prove that
\begin{equation}\label{eqn:int1}
\int_\Sigma d^4\hat x d^3u\,\left(\sqrt{g^L\, h}\; \delta_a A^a(x,y)\right)_{|\Sigma} = 
- \int_\Sigma d^4\hat x d^3u\,\left(\sqrt{g^L\, h}\; \big(\Gamma^{\delta\,p}{}_{pa}+S^p{}_{pa}\big) A^a\right)_{|\Sigma}
\end{equation}
and for n-homogeneous functions $A^a(x,y)$ that
\begin{equation}\label{eqn:int2}
\int_\Sigma d^4\hat x d^3u\,\left(\sqrt{g^L\, h}\; \bar\partial_a A^a(x,y)\right)_{|\Sigma} = 
- \int_\Sigma d^4\hat x d^3u\,\left(\sqrt{g^L\, h}\; \Big(g^{L\,pq}\bar\partial_a g^L_{pq} - \frac{2[4r+n-5]}{r(r-1)L}g^L_{ap}y^p\Big) A^a\right)_{|\Sigma}
\end{equation}

This overview completes the tools needed for field theories on Finsler spacetimes. In the following section we apply this newly
developed formalism to the case of generalized electrodynamics.

\section{Electrodynamics}\label{sec:ED}
By applying the technology that we have developed for field theory action integrals, we will formulate an explicit theory of generalized electrodynamics on Finsler spacetimes in this section. This theory reduces to standard electrodynamics if the Finsler spacetime is induced by a Lorentzian metric. Our key objective is the proof that the propagation of light indeed takes place on Finsler null geodesics. This claim, often found in the literature, will be confirmed here for the first time. The proof fundamentally relies on Theorem 2 of section~\ref{sec:geometry} that allows us to describe the differential geometry of the null structure. This in turn is a consequence of our definition of Finsler spacetimes. 

\subsection{Action and field equations}
Classical electrodynamics on a Lorentzian metric spacetime $(M,\tilde g)$ can be formulated in terms of an action for a one-form $\tilde A$ and a field strength two-form $\tilde F$ as
\begin{equation}\label{eqn:clED}
-\frac{1}{2} \int_M d^4x \sqrt{\tilde g}\;  \tilde g^{ab} \tilde g^{cd} \tilde F_{ac}\Big(\partial_b\tilde A_d-\partial_d\tilde A_b- \frac{1}{2}\tilde F_{bd}\Big) = \int_M d^4x \sqrt{\tilde g}\;\mathcal{L}(\tilde g_{ab}, \tilde A_a,\partial_a\tilde A_b,\tilde F_{ab})\,.
\end{equation}
The equations of motion are obtained from this action by variation with respect to $\tilde A$ and $\tilde F$:
\begin{equation}\label{eqn:cleq}
\nabla^{\tilde g}_b \tilde F^{ba} = 0 \,,\quad 
\tilde F_{ab} = \partial_a\tilde A_b-\partial_b\tilde A_a\,,
\end{equation}
where $\nabla^{\tilde g}$ denotes the Levi--Civita connection of the spacetime metric. The advantage of this formulation over the standard
action $-\frac{1}{4}\int d^4x\sqrt{\tilde g}\; \tilde F^{ab}\tilde F_{ab}$ lies in the fact that the relation $\tilde F=d\tilde A$, that
$\tilde A$ is a gauge potential, does not need to be imposed by hand.

In order to determine a generalized theory of electrodynamics on Finsler spacetimes $(M,L,F)$, we propose the following minimal extension principle:
\begin{itemize}
\item all equations of motion must be determined by the action;
\item all tensor fields of the theory on $(M,\tilde g)$ are lifted to zero-homogeneous tensor fields of the same type on $TM$, so that their components with respect to the Berwald bases are independent of the geometry $L$; 
\item the generalized action is obtained by using the scalar Lagrangian to contract the lifted fields and the metric $G$ of~(\ref{eqn:GTM});
\item the restriction of the result to $\Sigma$ is integrated as in (\ref{eqn:fact}); 
\item Lagrange multipliers are used to constrain all fields to their horizontal components.
\end{itemize}
This principle is designed to generate a unique extension of a given classical field theory. The last point is implemented to keep the same number of components of the physical fields as in the classical theory. Hence the extension essentially involves an additional dependence of the field components on the fibre coordinates which we can interpret in terms of observers in motion.

We now apply the minimal extension principle to the action~(\ref{eqn:clED}). Using the horizontal/vertical Berwald bases~(\ref{eqn:Berwald}), the fields $\tilde A$ and $\tilde F$ are lifted to\footnote{It should be clear from the context whether $F$ denotes the Finsler function or the field strength tensor.}
\begin{equation}
A = A_a(x,y) dx^a + A_{\bar a}(x,y) \delta y^a\,, \quad
F = \frac{1}{2}F_{ab}(x,y) dx^a\wedge dx^b +    F_{\bar a b}(x,y) \delta y^a\wedge dx^b   + \frac{1}{2}F_{\bar a\bar b}(x,y) \delta y^a\wedge \delta y^b\,.
\end{equation}
The forms $A$ and $F$ are required to be zero-homogeneous in the fibre coordinates; this implies the homogeneities zero for~$A_a,F_{ab}$, minus one for $A_{\bar a},F_{\bar a b}$ and minus two for $F_{\bar a\bar b}$. According to the  extension principle the generalized action then becomes
\begin{eqnarray}\label{eqn:eed}
S[A,F] & = &  \int d^4x d^3u \sqrt{g^L\,h}_{|\Sigma} \Big[-\frac{1}{2} G^{AB} G^{CD} F_{AC}\Big(\partial_BA_D-\partial_DA_B- \frac{1}{2}F_{BD}\Big) \nonumber \\ 
& & \hspace{3in}+ \lambda^{\bar a}A_{\bar a} + \lambda^{\bar a b}F_{\bar a b} + \lambda^{\bar a\bar b}F_{\bar a\bar b} \Big]_{|\Sigma}\,,
\end{eqnarray}
where the induced coordinates $(Z^A)=(x^a,y^a)$ and the corresponding partial derivatives are used before restricting to $\Sigma$. Observe the appearance of the Lagrange multipliers $\lambda^{\bar a}, \,\lambda^{\bar a b}$ and $\lambda^{\bar a\bar b}$ that kill the non-horizontal parts of $A$ and $F$ on-shell.

The variation of the generalized action with respect to $A$, $F$ and the Lagrange multipliers is technically straightforward; the calculation uses the Berwald bases and requires the integration by parts identities~(\ref{eqn:int1}) and~(\ref{eqn:int2}). Using the immediate constraints 
\begin{equation}
A_{\bar a} = 0 \,,\quad F_{\bar a b}=0\,,\quad F_{\bar a\bar b} =0 \,,\quad 
\end{equation}
we thus find the field equations
\begin{eqnarray}
F_{ab} & = & \delta_a A_b -\delta_b A_a\,,\label{eqn:FdA}\\
0 & = & g^{L\,ab} g^{L\,cd} \left(\nabla_a F_{bd} - S^p{}_{pa} F_{bd}\right),\label{eqn:nablaF}\\
\lambda^{\bar a b} &=& g^{L\,ap} g^{L\,bq} \bar\partial_p A_q\,,\quad \lambda^{\bar a}=\frac{1}{2}g^{L\,ab} g^{L\,cd}F_{ac}R^q{}_{bd}\,,
\quad \lambda^{\bar a \bar b}=0\,.\label{eqn:lam}
\end{eqnarray}
The field strength $F$ whose components are interpreted as electric and magnetic fields is gauge-invariant under the transformations 
\begin{equation}
A_a\mapsto A_a + B_a\,,\quad \delta_{[a}B_{b]}=0\,.
\end{equation}
These may change the solution for the Lagrange multipliers, but this has no physical relevance.

We emphasize that the field equations reduce to the standard Maxwell equations~(\ref{eqn:cleq}) in case the Finsler spacetime is induced by a Lorentzian metric and the fields only depend on the coordinates of the manifold $M$ but not on the fibre coordinates of $TM$.

\subsection{Propagation of light}
Any theory of electrodynamics determines the motion of light through the corresponding system of partial differential equations. Light trajectories are obtained in the geometric optical limit by studying the propagation of singularities of the electromagnetic fields. Since our field equations on Finsler spacetime are formulated over the tangent bundle, also the resulting singularity propagation will follow curves $\tau\mapsto (x(\tau),y(\tau))$ on the tangent bundle $TM$. Of these only the natural lifts $\tau\mapsto (x(\tau),\dot x(\tau))$ that arise from curves $\tau\mapsto x(\tau)$ on the manifold $M$ have an immediate interpretation as light trajectories. We will demonstrate the strong result that in the proposed extended electrodynamics~(\ref{eqn:eed}) all light trajectories are Finsler null geodesics.

In the following analysis we regard the components of the one-form $A$ as the fundamental variables. We insert (\ref{eqn:FdA}) into (\ref{eqn:nablaF}) to obtain the following system of linear second order partial differential equations
\begin{equation}
0 = g^{L\,a[b} g^{L\,d]c} \left(\nabla_a \nabla_b A_c - S^p{}_{pa} \nabla_b A_c\right).
\end{equation}
A solution of this system  for $A_a$ determines solutions for $F_{ab}$ and $\lambda^{\bar a b}$ according to our field equations~(\ref{eqn:FdA})--(\ref{eqn:lam}). Following standard methods for partial differential equations we now extract the principal symbol from the equations above. For this purpose we use the gauge condition $g^{L\,ab}\nabla_a A_b=0$ which generalizes the usual Lorentz gauge. Then the terms of highest derivative order can be written in the form
\begin{eqnarray}
0 &=& g^{L\,ab} \left(\partial_a\partial_b - 2 N^p{}_a \partial_b\bar\partial_p + N^p{}_aN^q{}_b\bar\partial_p\bar\partial_q \right)A_c + \dots\nonumber\\
&=& \delta^p_cP^{AB}  \partial_A\partial_BA_p +\dots
\end{eqnarray}
where the dots represent terms with less than two derivatives acting on the $A_a$. Following the definition of Dencker~\cite{Dencker}, and using the non-degeneracy of $g^L_{ab}$, one can check that the system is of real principal type. Hence, as Dencker shows, the propagation of singularities is governed by the Hamiltonian
\begin{equation}P(x,y,k,\bar k) = \frac{1}{2}P^{AB}k_Ak_B = \frac{1}{2} g^{L\,ab}k^H_ak^H_b\,,\quad k^H_a = k_a -N^p{}_a\bar k_p\,.
\end{equation}
More precisely, the singularities of the field $A$ propagate along the projection to $TM$ of the integral curves of the Hamiltonian vector field 
\begin{equation}
X_P = \partial_{k_a}P \partial_a + \partial_{\bar k_a}P \bar\partial_a - \partial_a P \partial_{k_a} - \bar\partial_a P \partial_{\bar k_a}
\end{equation}
that lie in the surface $P=0$. 

The integral curves $\tau\mapsto (x(\tau),y(\tau),k(\tau),\bar k(\tau))$ in $T^*TM$ of the Hamiltonian vector field $X_P$ are determined by the corresponding Hamiltonian equations
\begin{eqnarray}
\dot x^a &=& g^{L\,ab}k^H_b\,,\label{eqn:xdot}\\
\dot y^a &=& -g^{L\,pb}k^H_bN^a{}_p\,,\label{eqn:ydot}\\
\dot k_a &=& -\frac{1}{2}\partial_a g^{L\,pq}k^H_pk^H_q + g^{L\,pq}k^H_q\partial_a N^b{}_p\bar k_b\,,\\
\dot {\bar k}_a &=&  -\frac{1}{2}\bar \partial_a g^{L\,pq}k^H_pk^H_q + g^{L\,pq}k^H_q\bar \partial_a N^b{}_p\bar k_b\,.
\end{eqnarray}
and satisfy the constraint of lying in the surface $P=0$, i.e.,
\begin{equation}\label{eqn:M0}
g^{L\,ab}(x,y) k^H_a k^H_b =0\,.
\end{equation}
We can immediately conclude from (\ref{eqn:xdot}) and (\ref{eqn:ydot}) that the projection $\tau\mapsto(x(\tau),y(\tau))$ of these integral curves to~$TM$ satisfies $\dot y^a+N^a{}_p\dot x^p=0$, i.e., is horizontal. This yields the nice result that all singularities of the one-form $A$ propagate into directions that can be identified with tangent directions to the manifold. We also conclude from~(\ref{eqn:xdot}) and~(\ref{eqn:M0}) that  $\dot x^a$ is $g^L$-null. 

Light trajectories are a special case of the obtained propagation trajectories. They are natural lifts of some curve $\tau\mapsto x(\tau)$ in $M$ so that $y^a=\dot x^a$. Hence we conclude for light that
\begin{equation}
\ddot x^a + N^a{}_b(x,\dot x)\dot x^b =0\,,\quad g^{L\,ab}(x,\dot x)\dot x^a\dot x^b=0\,.
\end{equation}
Comparison of this result with (\ref{eqn:Fgeo}) and (\ref{eqn:gLL}) proves that all light trajectories that arise on Finsler spacetimes from the generalized theory of electrodynamics~(\ref{eqn:eed}) are Finsler null geodesics. This shows that the null structure of Finsler spacetimes is indeed related to the propagation of light.

\section{Discussion}\label{sec:disc}
In the motivation we have identified a number of conceptual issues that prevent a straightforward application of Finsler geometry to the description of spacetime. These involve imprecise definitions of Lorentzian Finsler structures, unclear notions of causality, and restricted constructions of field theory actions. Our results presented in this article demonstrate how to solve these problems:

We have developed a concise new definition for Finsler spacetimes $(M,L,F)$; this involves a fundamental geometry  function~$L$ on the tangent bundle that induces a Finsler function~$F$. From the properties of~$L$ we could prove that the construction of connections, covariant derivatives and curvature, which works perfectly on definite Finsler spaces, can be extended to our definition of Finsler spacetimes. In particular, these geometric objects were shown to be well-defined on the null structure. On the basis of this result, we could clarify the causal structure of Finsler spacetimes. With well-defined notions of timelike and lightlike vectors now available we can describe the trajectories of massive observers and point particles and of light as timelike or null Finsler geodesics, respectively. The timelike vectors form open convex cones with null boundary as in Lorentzian metric geometry. 

Further, we could deduce that light indeed must propagate along Finsler null geodesics. In order to obtain this result we developed a completely new formulation of well-defined field theory actions on Finsler spacetimes as integrals over the subbundle $\Sigma\subset TM$. These actions have no divergency problems and the corresponding field equations allow the full reconstruction of the physical fields. Our application of this method to a generalized theory of electrodynamics produced equations of motion that determine the propagation of light along Finsler null geodesics.

We conclude that Finsler spacetimes are consistent generalizations of Lorentzian metric manifolds that can be used as geometric spacetime backgrounds for physics. 

The following questions will be relevant for future research. The definition of observers on Finsler spacetimes needs to be made more precise. In particular this concerns the definition of spatial directions, the measurement of spatial length, and the group of transformations that relates different observers. Using the methods developed during this article it should be possible to formulate a gravity action for Finsler spacetimes that provides dynamics for the basic geometry function~$L$. As a guiding principle, these dynamics should be equivalent to the Einstein equations if the Finsler spacetime is induced by a a Lorentzian metric. 
Further, it would be interesting to study how satisfactorily the technology we have developed can be applied to crystal media described by bimetric Finsler functions as in the second example presented in the examples' section. Is it possible to deduce the effects observed in these crystals from our generalized electrodynamics, and can we learn something about the interpretation of the tangent direction dependence of the fields?

Besides these questions that concern the physical interpretation and application of Finsler spacetimes there is also the mathematical question about the relation between Finsler spacetimes and other generalized geometric backgrounds for physics. For example, is there a connection to  the backgrounds based on hyperbolic polynomials on the cotangent bundle of a manifold~\cite{Raetzel:2010je}? The examples discussed in this article are not only Finsler spacetimes but also belong to that class. It would be interesting to investigate whether there is an overlap of spacetimes consistent with the different definitions. We note the fact that bimetric backgrounds defined by two Lorentzian metrics with timelike cones that intersect only in the origin are excluded both as Finsler spacetimes and as hyperbolic polynomials. 
 
Using the techniques presented in this article we already study observers and gravity on Finsler spacetimes. We expect to report on further results soon.

\acknowledgments
The authors are happy to thank Claudio Dappiaggi, Manuel Hohmann, Matthias Lange and Gunnar Preiss for very fruitful discussions, and Ioan Bucataru for constructive comments. They  gratefully acknowledge full financial support from the German Research Foundation DFG through the Emmy Noether grant WO~1447/1-1.


\end{document}